\documentclass[12pt]{iopart}

\usepackage{xcolor}
\usepackage{svg}
\usepackage{relsize}
\usepackage{graphicx}

\newcommand{\futurenote}[1]{}
\newcommand{\cpchanges}[1]{#1}
\newcommand{\tcb}[1]{#1}
\begin{document}

\title{A finite-difference model for intense light
interactions with dielectrics in the
ultrafast ionization regime}

\author{Julia Apportin$^1$, Christian Peltz$^1$, Pavel Polynkin$^2$, Misha Ivanov$^3$, Thomas Fennel$^1$,  Anton Husakou$^3*$}

\address{$^1$University of Rostock, Universit\"atsplatz 1, D-18055 Rostock, Germany}
\address{$^2$University of Arizona, Tucson, AZ 85721, USA}
\address{$^3$Max Born Institute, Max Born Str. 2a, D-12489 Berlin, Germany}

\ead{gusakov@mbi-berlin.de}
\vspace{10pt}
\begin{indented}
\item[]March 2026
\end{indented}

\begin{abstract}
We present a computationally efficient model that describes
the interaction of intense, ultrashort infrared laser pulses with transparent
materials in the strong ionization regime. The model is augmented with a
detailed self-consistent description of the local response due to ionization
and collisional plasma dynamics. It incorporates the direct solution of
Maxwell’s equations without approximations and rigorous boundary
conditions for the input pulse, allowing us to study the ultrafast formation of
over-critical nanoscaled plasmas in dielectric materials under the influence
of intense tightly focused laser pulses. We perform a scan of the parameter space, find unexpected optima regimes for different experientially relevant parameters, and explain these maxima based on spatiotemporal dynamics. 
\end{abstract}

\section{Introduction}
The interaction of intense, ultrashort infrared laser pulses with transparent materials addresses fundamental questions of light-matter interaction and is of key importance for technological applications. Relevant fundamental problems include attosecond and few-femtosecond dynamics of strongly driven electrons in valence and conduction bands \cite{o3}, the emergence of effective band structures in the regime of strongly driven Bloch oscillations \cite{o4}, attosecond polarization and currents in
light-driven optical elements for lightwave electronics \cite{o5}, the attosecond dynamics of material breakdown \cite{o8}, the extreme conditions of pressure and temperature created in bulk by intense laser pulses, and the appearance of new material phases under such high-pressure, high-temperature conditions \cite{o9}. These fundamental questions are of key importance for technological applications, such as
 laser machining of surfaces and bulk \cite{o10,o11}, medical application e.g. laser eye surgery with femtosecond lasers \cite{o12}, etc.

There is a plethora of physical effects that have to be incorporated in the modeling the
material response \cite{o20} in the high-intensity tight-focusing regime, such as plasma formation
and dynamics, energy dissipation by collisions, changing velocity distributions of the electrons,
modification of refractive index etc. The most accurate and efficient description
of all these effects, in combination with light propagation, can be achieved by first-principle methods such as MicPIC
(Microscopic Particle-in-Cell) approach, e.g. \cite{o7,o26}. While offering a detailed insight into the microscopic dynamics of plasma formation and its interaction with the
laser field, the MicPIC method is currently unable to perform the simulations at the macroscopic (several micrometers and above) spatial dimensions. Thus, the development of simplified
material response models remains essential.

 From the theoretical perspective, one of the key challenges in modeling
the ultrafast strong-field response of a bulk dielectric to the propagation of intense, tightly focused light is the ultrafast change in the refractive properties of the material. An over-critical plasma can arise in just a couple of optical cycles, starting to reflect the incident light. Under such conditions none of the standard approximations such as the slowly evolving wave approximation \cite{o22}, unidirectional pulse propagation \cite{o23}, the paraxial approximation, etc. are applicable; also,  azimuthal symmetry is broken by the polarization of the tightly-focused light. Full solution of Maxwell’s equations is required, in addition, rapid ionization and the ultrafast change of the charge density leads to generation of harmonics, due to e.g. the Brunel mechanism \cite{o25} which imposes additional demand on the spatio-temporal grid for the Maxwell’s equations solver.

The nonlinear propagation of pulsed beams through dielectrics was extensively studied in
the past, including the effects of Kerr nonlinearity, dispersion, photoionization, and avalanche
ionization \cite{o13,o14,o15,o16,o17,o18,o19,o20,o21}.
In the recent years, several groups have developed models which adequately include most of the abovementioned effects. As two prominent examples, we mention Ref. \tcb{\cite{ex1} and \cite{ex2}}, which augment full solution of Maxwell equations in the FDTD framework with detailed microscopic description of the material response. However, in both cases there are some drawbacks which can influence the accuracy of the simulation. For example, in \cite{ex1} the collision rate is independent on the average (drift) velocity of the electrons, which does not correspond to the realistic physical picture, and the mobility terms are introduced ad-hoc without microscopic justification. In a similar vein, in \cite{ex2} avalanche and elastic-collision rates are formulated in terms of temperature which is expressed in terms of heat capacity ratio, which is, however, irrelevant for the microscopic properties.

Motivation of the present works goes beyond the abovementioned drawbacks of the existing models. The dependence of the key parameters of the light-matter interaction, such as the maximum deposited energy and the volume of over-critical plasma, on the excitation parameters (such as numerical aperture of the focusing an pulse duration) remains a topic of the ongoing debate. For example, for the fixed pulse energy it is often considered favorable to take the pulse duration as short as possible an to focus it as tightly as possible. Here we perform, based on our \tcb{rigorous} model, an extensive scan of the input parameters. We predict the existence of optima in the input parameter space which correspond to relatively long (hundreds of femtoseconds) pulses and provide the maximum deposited pulse energy. In a similar vein, we show that for the maximum overcritical plasma volume, moderate focusing provides better results than the tight focusing. We physically explain these counterintuitive findings by spatiotemporal dynamics of the pulse propagation and plasma excitation, as detailed below.

The paper is organized as follows. In Section 2, we systematically derive the local dynamics equations for the drift velocity and the dispersion of the electron velocity (temperature) based on features of different processes relevant for photoionization, collisions and electron acceleration. We establish boundary conditions which accurately take into account the fully vectorial nature of the field, \tcb{absence} of the \tcb{input} evanescent waves, pulse reflection at the boundary of the sample, as well as numerical group velocity dispersion. In Section 3, the model is used to systematically scan the parameter space and to evaluate the parameter dependence of several key quantities, such as deposited energy, maximum achieved intensity, and volume of overcritical plasma.

\section{Numerical model}\label{sec:model}
\subsection{Maxwell equations}
The most general way to describe non-linear pulse propagation is the direct solution of Maxwells equations, which is numerically very demanding. Common approaches to reduce the computational effort include (i) the neglect of backpropagating waves, which is allowed for weakly inhomogeneous media and yields a first order propagation equation \cite{geissler_1999}, (ii) a paraxial approximation (weak focusing) that yields the Gaussian beam solution and allows a split step description by decoupling spatial and temporal effects\cite{splitstepref}, and the scalar approximation that neglects the coupling of field components~\cite{tuemmler26}. None of the above approximations are applicable to the considered scenario of tightly focused ultra-short laser pulses, \tcb{so} that a rigorous treatment starting from first principles is required.

We depart from the macroscopic formulation of Maxwells equations for the real-valued electric $\mathbf{E}$ and magnetic $\mathbf{H}$ fields
\begin{eqnarray}
            \frac{\partial\mathbf{E}}{\partial t}&=+\frac{1}{\varepsilon_0}(\nabla \times \mathbf{H}-\mathbf{J}) \\
        \frac{\partial\mathbf{H}}{\partial t}&=-\frac{1}{\mu_0}(\nabla \times \mathbf{E}) \label{eq:maxwell4}.
\end{eqnarray}
Here the current density $\mathbf{J}$ contains the linear and nonlinear material response with contributions from bound and quasi-free electrons,
as well as the currents associated with the promotion of electrons to the conduction band via strong field ionization:
\begin{equation}
    \mathbf{J} = \mathbf{J}_{Bound} +\mathbf{J}_{SFI} + \mathbf{J}_{Free}.
\end{equation}
The bound electron contribution, which governs the neutral bulk response, is given by the corresponding polarization $\mathbf{P}_{B}$ via
\begin{equation}
    \mathbf{J}_{Bound} = \frac{\partial \mathbf{P}_B}{\partial t}
\end{equation}
and can be described by a collection of Drude-Lorentz type oscillators
\begin{equation}
    \mathbf{P}_B = (1-\rho) \sum_i \mathbf{P}_{B,i}.
\end{equation}
Here $\rho$ denotes the local fraction of ionized entities and the prefactor $(1-\rho)$ scales the bound electron contribution accordingly, \tcb{whereby we neglect the typically weak response of bound electrons on lower-lying energy states}.
For the individual oscillators we employ the extended Drude-Lorentz model described in Ref.~\cite{varin1}
\begin{equation}
    \frac{\partial^2\mathbf{P}_{B,j}}{\partial t^2}+\gamma_j\frac{\partial \mathbf{P}_{B,j}}{\partial t}+\omega^2_{E,j}\mathbf{P}_{B,j}=\epsilon_0 \omega^2_{E,j} \left(\bar{\chi}^{(1)}_j\mathbf{E}+\bar{\chi}^{(3)}_j|\mathbf{E}|^2\mathbf{E}\right)
\end{equation}
which includes both linear and Kerr-type nonlinear terms. The parameters $\omega_{E,j}$, $\gamma_j$ are related to the resonance frequency and the damping of resonance $j$ while $\bar{\chi}^{(1)}_j$ and $\bar{\chi}^{(3)}_j$ determine the oscillator strength of the linear and nonlinear contributions, respectively.

For the treatment of photoionization, we substitute the solid-state material by a dense ensemble of entities, such as molecules and atoms, with each entity being photo-ionized separately. This is an often used and standard approximation in the modeling of strong-field interaction with dielectric materials. The ionization current $\mathbf{J}_{SFI}$ arises from the displacement of the electron from the ionic core immediately after the photoionization. It describes the transfer of bandgap energy $E_g$ per photoionization event from the electromagnetic pulse to the electron system. It is given by 
\begin{equation}
    \mathbf{J}_{SFI} =-n_0(1-\rho)E_g\frac{\gamma_{SFI}\mathbf{E}}{|\mathbf{E}|^2}
\end{equation}
with the density of neutral entities $n_0$ and the field ionization rate $\gamma_{SFI}$. 

Finally, the contribution of quasi-free electrons, i.e. electrons already promoted to the conduction band by field ionization or collisional ionization, needs to be taken into account. The resulting material current is determined by the their respective average drift velocity $\mathbf{v}_D$ via
\begin{equation}
    \mathbf{J}_{Free} = - \rho e n_0 \mathbf{v}_D.
\end{equation}
The corresponding physical model for $v_D$ is the core-component of our description and will be detailed in the following subsections.

\subsection{Model of collisions}
In order to describe the dynamics of the free-electron current $\mathbf{J}_{Free}$, we need to consider the microscopic processes of electron collisions as well as photoionization and avalanche ionization processes.
In the framework of our model, the following processes are taken into account: photoionization of neutral entities (molecules/atoms), avalanche ionization by free electrons colliding with neutral atoms, elastic collisions of free electrons with neutrals and ions, as well as electron-electron collisions. 

It is important to note that the expressions presented below provide a unifying platform which allow to incorporate different types of light-electron, electron-electron and electron-ion interactions. For each interaction, just three quantities need to be defined: a rate of the process, a change of the electron drift velocity, and a change of the electron velocity dispersion. This approach provides a systematic paradigm to the formulation of the material equations and allows to avoid double-counting of the elementary physical processes through first-principle and phenomenological terms.

In the framework of the proposed model, \cpchanges{we mainly study} the early stages of the processes associated with strong field pulse propagation in dielectrics. We \cpchanges{typically employ} laser pulses with durations below $\sim$500 fs\cpchanges{, where some of the slower} processes, such as recombination as well as the temperature transfer to the lattice\cpchanges{, can be safely neglected.} \cpchanges{Note that, as discussed later, we also consider laser pulses with picosecond durations in this study to illustrate trends beyond the region of applicability; however, as shown by the results of the parameter space scan with fixed pulse energy, total ionization is significantly reduced in these cases and the above-mentioned slow processes do not play any significant role.} Also, we do not consider the processes where a fast electron collides with an ion and changes its electronic state, since the relative ionization is typically below 30\% in our case and such collisions constitute a minor contribution. By the same reason, the elastic collisions with ions and neutrals are combined in the same process. In this subsection, we consider each of the above-mentioned processes separately and derive the change of the electron average velocity and temperature, as defined below.

We begin with the microscopic model of collisions. Let $n_0$ be the neutral entities number density, $n_e$ the number density of electrons in the conduction band, $\rho = n_e/n_0$ the relative ionization degree, and $(1-\rho)n_0$ the number density of neutrals left. Currently, double ionization is ignored, but the formalism can be extended in a straightforward way to include it.

We describe the motion of quasi-free electrons by two quantities which are functions of time and all spatial coordinates: vectorial drift velocity $\mathbf{v}_d$ and scalar thermal velocity ${v_T}$. The average (drift) velocity $\langle \mathbf{v} \rangle = \mathbf{v}_d$ is associated with the current in the medium and is imparted by the laser field. In addition, electrons have a dispersion of velocities, which corresponds to the electron temperature and is
characterized by $v_T$ given by $\langle \mathbf{v}^2\rangle =v^2_T+\mathbf{v}^2_d$. In doing this, we assume that the random component of velocity characterized by $v_T$ is isotropic, and due to fast thermalization within the electron cloud, the distribution of the velocities is Maxwellian.

Elastic and inelastic collisions of electrons with the lattice are described via corresponding rates $\gamma_{el}$ and $\gamma_{in}$. In our case, the dominant inelastic channel is the avalanche channel. Elastic collision rates can be different for neutrals and ions, with the average collision rate given by
$\gamma_{el}\to\gamma_{el}^{(i)}\rho+\gamma_{el}^{(n)}(1-\rho)$.  In the present implementation, the inelastic collision rate $\gamma_{in}$ only includes collisions with neutrals.

The number density of electrons not experiencing any collisions in a time interval $\Delta t$ is $n_1 = n_e(1-\Delta t(\gamma_{in}+\gamma_{el}))$. The corresponding velocities of this group remain unchanged:
\begin{eqnarray}
&&\langle \mathbf{v}_1 \rangle=\mathbf{v}_d,\nonumber\\
&&\langle \mathbf{v}_1^2 \rangle=v_T^2+\mathbf{v}_d^2,\nonumber\\
&&n_1 = n_e(1-\Delta t(\gamma_{in}+\gamma_{el}))
\end{eqnarray}

The number density of electrons ionized across the bandgap after inelastic collisions is $n_2 = n_e\gamma_{in}\Delta t$. We assume that these electrons are freed with zero velocity and energy, i.e. $\langle \mathbf{v}_2^2\rangle=\langle \mathbf{v}_2\rangle=0$ at the moment of their production:
\begin{eqnarray}
&&\langle \mathbf{v}_2 \rangle=0,\nonumber\\
&&\langle \mathbf{v}_2^2 \rangle=0,\nonumber\\
&&n_2 = n_e\Delta t \gamma_{in}.
\end{eqnarray}

The number density of quasi-free electrons that have experienced inelastic collisions is $n_3 = n_e\Delta t \gamma_{in}$. We assume that after collisions these electrons have zero average velocity, and that their energy decreases by the energy $E_g$ lost for collisional excitation. Hence, immediately after
the inelastic collision, we write for this group of electrons
\begin{eqnarray}
&&\langle \mathbf{v}_3 \rangle=0,\nonumber\\
&&\langle \mathbf{v}_3^2 \rangle=v_T^2+\mathbf{v}_d^2-2E_g/m,\nonumber\\
&&n_3 = n_e\Delta t\gamma_{in}.
\end{eqnarray}

For elastic collisions \tcb{with neutrals and ions}, the energy is conserved but the direction of electron velocity is randomized, leading to zero average velocity (isotropic cross-section),
\begin{eqnarray}
&&\langle \mathbf{v}_4 \rangle=0,\nonumber\\
&&\langle \mathbf{v}_4^2 \rangle=v_T^2+\mathbf{v}_d^2,\nonumber\\
&&n_4 = n_e\Delta t\gamma_{el}.
\end{eqnarray}

For elastic electron-electron collisions, both average velocity of the colliding pair and its total energy do not change during collision:
\begin{eqnarray}
&&\langle \mathbf{v}_5 \rangle=0,\nonumber\\
&&\langle \mathbf{v}_5^2 \rangle=v_T^2+\mathbf{v}_d^2.\nonumber\\
\end{eqnarray}

Finally, the number density of electrons which become quasi-free by photoionization is $n_6 = (1-\rho)n_e\Delta t\gamma_{\mathrm{SFI}}$, where $\gamma_{\mathrm{SFI}}$ is the photoionization rate, and we assume that these electrons appear with zero velocity:
\begin{eqnarray}
&&\langle \mathbf{v}_6 \rangle=0,\nonumber\\
&&\langle \mathbf{v}_6^2 \rangle=0,\nonumber\\
&&n_6 = (n_0-n_e)\Delta t\gamma_{\mathrm{SFI}}.
\end{eqnarray}

\subsection{Models for collision and ionization rates}

Next, we introduce models for elastic and inelastic collision rates. As noted above, we assume that the electrons have a Maxwellian distribution over the velocity $\mathbf{v}$ given by $M(\mathbf{v}_d,v_T;\mathbf{v})$, which depends on the drift velocity $\mathbf{v}_d$, as well as the thermal velocity $v_T$. 

Both elastic- and inelastic-collision rates are modeled as
\begin{equation}
    \gamma_{el,in}=n_{el,in}\int d^3v|\mathbf{v}|\sigma_{el,in}(|\mathbf{v}|)M(\mathbf{v}_d,v_T;\mathbf{v}),
    \label{eq:collision_rate2}
\end{equation}
i.e. they are proportional to the corresponding (elastic or inelastic) angle-average transport cross section $\sigma_{e,i}(|\mathbf{v}|)$, and the absolute velocity of the particle $|\mathbf{v}|$. The number density $n_{el,in}$ in front takes into account that in our model only remaining neutrals can be targeted by inelastic collisions ($n_{in}=n_0-n_e$) while elastic collisions apply to all entities ($n_{el}=n_0$).

In the present implementation, the elastic cross sections are modeled as $\sigma(|\mathbf{v}|) = \sigma/[1+|\mathbf{v}|^2/v^2_0]$, with $v_0$ a fitting parameter \futurenote{(currently set to one atomic unit)}; the inelastic collisions are modeled using the Lotz formulas \cite{o27}. Note that electron-electron collisions do not directly influence either $\mathbf{v}_d$ or $v_T$, however, as a result of such collisions the electron velocity distribution becomes
Maxwellian. We stress that both collision rates strongly depend on the electron average velocity $\mathbf{v}_d$ and temperature $T$ and are therefore time- and space-dependent, since velocity and temperature change in time and space very significantly. The velocity integrals were pre-calculated and the cross-sections for elastic and inelastic collisions were tabulated as functions of both $\mathbf{v}_d$ and $v_T$, to be efficiently used during the numerical simulation of propagation.

The field ionization rate $\gamma_{SFI}$ is modeled on the sub-cycle time-scale by ADK ionization formula \cite{new2}, with a pre-exponential factor which was fit to reproduce experimentally measured \cite{o29} cycle-averaged ionization rates in the material (silica).

Next, we derive the dynamics equations for the current and the temperature that will determine the dynamics of the parameters
$\mathbf{v}_d$ and $v_T$.

\subsection{Equations for current and temperature}
With the microscopic model of collisions in place, we can compute the averages and their
changes after a time step $\Delta t$. We start with the average velocity after collisions,
\begin{equation}
    \langle \mathbf{v} \rangle=\frac1{\sum_in_i}\sum_in_i\langle \mathbf{v}_i\rangle
\end{equation}
where
\begin{equation}
\sum_i n_i = n_e + \Delta t \left(n_e\gamma_{in} +  (n_0-n_e)\gamma_{\mathrm{SFI}}\right)
\end{equation}
Here \tcb{$\gamma_{\mathrm{SFI}}$} is the ionization rate which depends on the instantaneous field, and $(n_0-n_e)$ describes the depletion of neutrals. The \tcb{$\Delta t n_e\gamma_{in}$} term \cpchanges{describes the inelastic collisions. Keep in mind that the depletion of neutrals reduces the number of valid partners for inelastic collisions, an effect that is already included in the definition of the rates $\gamma_{in}$, c.f. eq.~\ref{eq:collision_rate2}.}

Substituting expressions for $n_i$ and $\mathbf{v}_i$, which accounts for the increased number density of the electrons, and keeping in mind that $\Delta t \to 0$ and hence $1/(1+A\Delta t)= 1-A\Delta t$, we get
\begin{equation}
\langle \mathbf{v}\rangle = \mathbf{v}_d - \Delta t \mathbf{v}_d\left(2\gamma_{in} + \gamma_{el} + \frac{n_0 - n_e}{n_e}\gamma_{\mathrm{SFI}}\right)
\end{equation}
where higher order terms in $\Delta t$ have been neglected. For the current density $\mathbf{J}_{free}$, however, the expression changes:
\begin{equation}
    \mathbf{J}_{free}=-e\left[\sum_in_i\right]\langle \mathbf{v}\rangle=-e\sum_in_i\langle \mathbf{v}_i \rangle=-en_e\mathbf{v}_d[1-\Delta t(\gamma_{in}+\gamma_{el})].
\end{equation}
The factor 2 in front of $\gamma_{in}$ disappears: the loss of average velocity is partially offset by the increase in the number of electrons. Thus, the contribution of collisions to the differential equation for the current is, as expected,
\begin{equation}
    \frac{\partial \mathbf{J}}{\partial t}|_{col.}=-(\gamma_{el}+\gamma_{in})\mathbf{J}. \label{eq:prop_j_col}
\end{equation}
Of course, there \tcb{is} also \tcb {a} term due to the electric field, so the full equation reads
\begin{equation}
    \frac{\partial \mathbf{J}}{\partial t}=-(\gamma_{el}+\gamma_{in})\mathbf{J}+n_ee^2\mathbf{E}/m_e. \label{eq:prop_j_free}
\end{equation}
Note that we ignore the effect of the magnetic field component (the Lorentz force) on the electrons since we consider non-relativistic motion.

Next, we derive the equation for the thermal velocity which is defined as
\begin{equation}
    \langle v_T^2\rangle=\langle \mathbf{v}^2 \rangle-\langle \mathbf{v} \rangle^2.
\end{equation}
\cpchanges{and trivially connected to the electron temperature via $T=m_e/k_BT v_T^2$.} First, we compute $\langle \mathbf{v}^2 \rangle$,
\begin{equation}
    \langle \mathbf{v}^2 \rangle=\frac1{\sum_in_i}\sum_in_i\langle \mathbf{v}_i^2\rangle.
\end{equation}
and, substituting all expressions from the microscopic model above, we get
\begin{equation}
\langle \mathbf{v}^2 \rangle = v_T^2 + \mathbf{v}_d^2 - \Delta t \left[ \gamma_{\mathrm{in}} (v_T^2 + \mathbf{v}^2_d + 2E_g/m_e) + \frac{n_0 - n_e}{n_e}\gamma_{\mathrm{SFI}}(v_T^2 +\mathbf{v}^2_d) \right]
\end{equation}
Next, we compute $\langle \mathbf{v} \rangle^2$, using the already derived expression for $\langle \mathbf{v} \rangle$ and again omitting higher order terms
\begin{equation}
    \langle \mathbf{v} \rangle^2=\mathbf{v}_d^2\left(1-\Delta t[4\gamma_{in}+2\gamma_{el}+2\frac{n_0-n_e}{n_e}\gamma_{SFI}]\right)
\end{equation}
which finally yields the change $\Delta v_T$ in thermal velocity  $v_T$, and the corresponding dynamics equation
\begin{equation}
    \frac{\partial}{\partial t}\left(v_t^2\right) = - \gamma_{in} 2E_g/m_e -\gamma_{in} (v_t^2-\mathbf{v}_d^2)  - \frac{n_0 - n_e}{n_e}\gamma_{\mathrm{SFI}}(v_t^2 - \mathbf{v}_d^2)+ 2(\gamma_{in} + \gamma_{el})\mathbf{v}_d^2
\end{equation}
This completes the closed system of equations for the local response of the medium. The resulting system of equation was solved using the standard FDTD leapfrog scheme, with perfectly-matched-layer boundary conditions. Note that this model, unlike many unidirectional models, satisfies the Kramers-Kronig relations by design, since field values are updated in time, guaranteeing causality.

\subsection{Model of tight focus}
Finally, we need to incorporate the external laser pulses into our model. Since we aim at analyzing scenarios involving tightly focused beams, our conditions will be well outside the range of validity of the standard paraxial approximation, i.e. the typically applied Gaussian Beam solution is no longer valid. We therefore adopt a more general approach that is based on the angular spectrum propagation method~\cite{HechtNovotny}. 

Specifically, we begin with a Fourier representation that connects the electric and magnetic fields at arbitrary position $\mathbf{r}$ and time $t$ with the angular spectrum of a two-dimensional transverse field slice at $z=0$ representing the transverse laser profile. This method not only allows us to accurately model essentially arbitrary laser profiles (as long as they obey Maxwell's equations), but also effectively reformulates the propagation problem in terms of plane waves. This substantially simplifies the numerical treatment as each of the plane waves propagates independently within linear media. For homogeneous linear media, the resulting representations of the electric and magnetic fields are given by
\begin{equation}
\mathbf{E}(\mathbf{r},t) = \frac{1}{(2 \pi)^{3/2}} \int\!\!\int\!\!\int_{-\infty}^\infty \mathrm{d}{\omega} \mathrm{d}{k_x} \mathrm{d}{k_y} \mathbf{\hat{E}}(k_x,k_y,z = 0,\omega) e^{i(k_x x + k_y y + k_zz - \omega t)}
\label{eq:focus_E}
\end{equation}
and
\begin{equation}
\mathbf{H}(\mathbf{r},t) = \frac{1}{(2 \pi)^{3/2}} \int\!\!\int\!\!\int_{-\infty}^\infty \mathrm{d}{\omega} \mathrm{d}{k_x} \mathrm{d}{k_y} \frac{i}{\omega \mu_0} \mathbf{k} \times \mathbf{\hat{E}}(k_x,k_y,z = 0,\omega) e^{i(k_x x + k_y y + k_zz - \omega t)},
\label{eq:focus_H}
\end{equation}
respectively. The angular spectrum $\hat{\mathbf{E}}(k_x,k_y,z = 0,\omega)$ describes the spectral amplitudes and polarization direction of the plane waves. It can be factored into two parts, a complex amplitude $\hat{\mathcal{E}}$ and a polarization vector $\hat{\mathbf{e}}_p$ such that
\begin{equation}
    \hat{\mathbf{E}}(k_x,k_y,z = 0,\omega) = \hat{\mathcal{E}}(k_x,k_y,z = 0,\omega) \, \hat{\mathbf{e}}_p(k_x,k_y,z = 0,\omega).
\end{equation}
We now connect the complex amplitude in Fourier-space to a transverse laser profile $\mathcal{E}$ at z=0 via
\begin{equation}
\hat{\mathbf{\mathcal{E}}}(k_x,k_y,z = 0,\omega) = \frac{1}{(2 \pi)^{3/2}}\int\!\!\int\!\!\int_{-\infty}^\infty \mathrm{d}{t} \mathrm{d}{x}\mathrm{d}{y} \mathbf{\mathcal{E}}(x,y,z = 0,t) e^{-i(k_x x + k_y y - \omega t)}.
\end{equation}
In this work we employ standard Gaussian profiles for the spatial and the temporal electric field profile
\begin{equation}
\mathcal{E}(x,y,z=0,t) = E_0 e^{-\frac{x^2 + y^2}{w_0^2}} e^{-2\ln(2) \frac{t^2}{\tau^2}}\left[\cos{(\omega_c t + \varphi_c)}-\frac{t}{\omega_c\tau^2}\sin{(\omega_c t + \varphi_c)}\right]
\label{eq:transverse_E_profile}
\end{equation}
with field amplitude $E_0$, beam width $w_0$, pulse length $\tau$, central laser frequency $\omega_c$ and carrier envelope phase $\varphi_c$.
The additional sinusoidal carrier term in eq.~\ref{eq:transverse_E_profile} ensures physically meaningful vector potentials, i.e. such that $\int_{-\infty}^\infty E dt=0$ holds also for very short pulses.
 \cpchanges{Note that the resulting distribution contains evanescent waves for ($k_x^2+k_y^2>\omega_c^2/c^2$), which we omit by limiting the integration to propagating waves.}
 
Finally the polarization vector $\hat{\mathbf{e}}_p$ can be determined. In order to yield solutions that obey Maxwell's equations, it has to satisfy
\begin{equation}
    \mathbf{k} \cdot \hat{\mathbf{E}} = \mathbf{k} \cdot \mathbf{e}_p\hat{\mathcal{E}} = 0,
\end{equation}
for each plane wave component, i.e. the polarization has to be perpendicular to the propagation direction. Considering x as the main polarization direction ($E_y=0$), leads to the following requirement
\begin{equation}
\mathbf{e}_p(k_x,k_y,\omega)=\frac{(k_z \mathbf{e}_x-k_x \mathbf{e}_z)}{\sqrt{k_x^2+k_z^2}}.
\end{equation}

From an implementation point of view individual plane waves enter and exit the computational domain according to the Total-Field/Scattered-Field (TFSF) method, in which electric and magnetic fields are decomposed into incident and scattered components. The incident fields are explicitly introduced (added) or subtracted (removed) at the TFSF boundaries, ensuring that no incident field is present outside the central Total-Field region~\cite{o24}. Accurate removal of these fields is critical, as residual fields at the boundary would otherwise act as artificial radiation sources. Achieving this clean field removal requires precise tracking of their temporal and spatial evolution (see eqs. \ref{eq:focus_E}, \ref{eq:focus_H}), which in turn requires accurate knowledge of the materials dispersion relation $k_z(k_x,k_y,\omega)$. A detailed description of the incident field treatment, including numerical dispersion corrections and the handling of an additional interface via numerically corrected Fresnel equations, is beyond the scope of this work and will be discussed elsewhere.

\section{Results}
In the next step we use our model to explore the modification of solid dielectric material under illumination with a tightly focused near infrared laser pulse. Specifically, we will investigate the dynamics of the energy deposition as well as the plasma generation near the focus as a function of the focal spot size and the pulse duration. A main goal of our analysis is to clarify the importance of the dynamical feedback between transient optical properties and plasma dynamics for the physically correct description of the associated laser-matter interaction.

We consider excitation of fused silica with a focused 800-nm laser pulse with Gaussian spatial and temporal profile at a fixed total pulse energy of 30 nJ. The focal point inside the medium is positioned 13 $\mu$m behind the surface of the fused-silica sample. The linear dispersion and the perturbative nonlinear response of the material due to a Kerr nonlinearity is described by a set of nonlinear driven Drude-Lorentz Oscillators with parameters taken from \cite{varin1}. For the observables of interest in our study the contribution of Raman terms to the nonlinearity is of minor importance and thus neglected. The instantaneous ionization probability is modeled using the "PPT-Rate" \cite{PPT_Rate} assuming a bandgap of 7.7 eV \cite{Jürgens_2020}. Electron impact ionization due to inelastic collisions is described with Lotz-cross-sections \cite{crossection_params} \futurenote{(can't reproduce elastic collision parameter from that paper used $v_0 = 1.4114\times10^6m/s, \sigma_0 = 30.738\AA{}^2$, used lotz cressection formular for $E\approx E_g$ with $E_g = 7.7eV, a = 4.5e^2, q = 2.1, b=c=0.6$, "a" and "q" are same for most elements; "b" and "c" is in that range for most elements, but not known exactly)}while the elastic electron-ion collisions are modeled with cross-section according the approach described above\futurenote{, add specific numbers for lotz and e-ion colls somewhere}.

For our systematic analysis we considered 7 pulse durations in the range from 3 to 3000 fs and 3 focal spot sizes ($w_0$) from 400 to 2191 nm. For each point of the resulting parameter space, we performed two simulations to highlight the significance of the dynamical feedback effects on ionization, absorption and propagation. One of the simulations reflects the case with full feedback, i.e. the complete modeling according to the approach described in section~\ref{sec:model} with plasma and Kerr terms. The second simulation describes a purely linear pulse propagation, without plasma feedback and Kerr nonlinearity. To this end the plasma contribution to the optical properties via the associated currents $\mathbf{J}_{Free}$ and $\mathbf{J}_{SFI}$ as well as the Kerr term in $\mathbf{J}_{bound}$ are neglected. As a result the plasma generation and heating is driven by the field predicted by a hypothetical purely linear dispersive pulse propagation. In consequence effects like the removal of the field energy spent for ionization as well as emerging reflection due to a plasma mirror are not accounted for.

\begin{figure}
    \centering
    \small
    \includegraphics[width=1.00\linewidth]{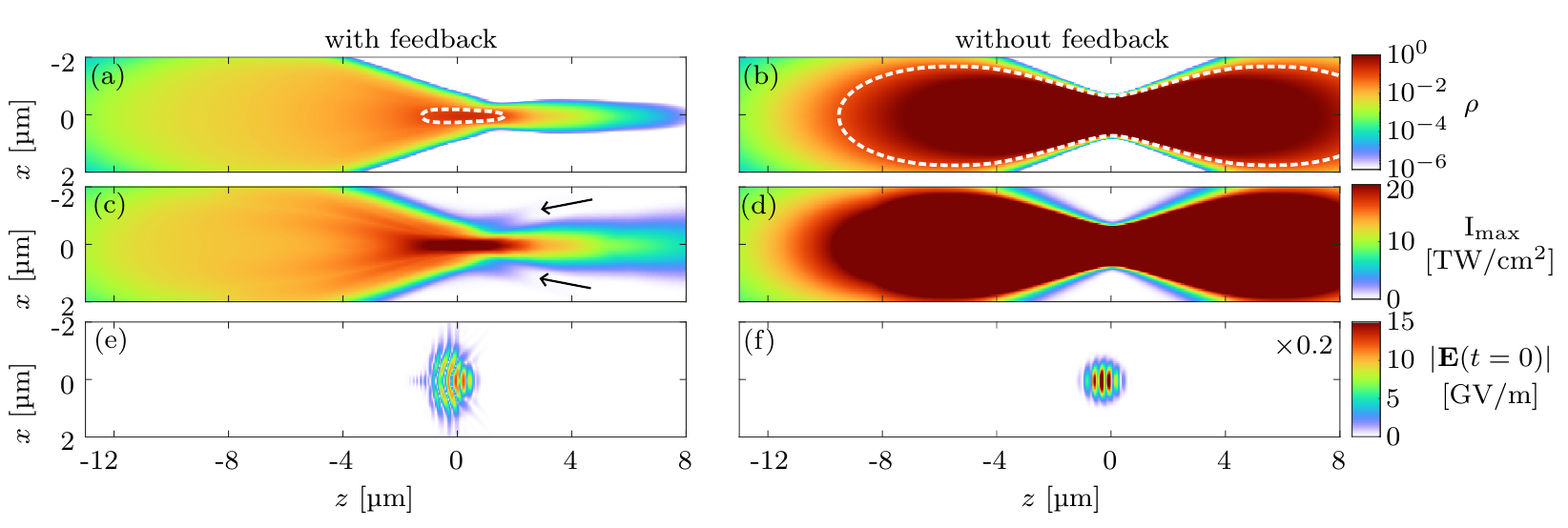}
    \normalsize
    \caption{Comparison of pulse propagation results with (left) and without (right) nonlinear feedback for the shortest pulse ($3\,\mathrm{fs}$) and strongest focusing ($w_0=400\,\mathrm{nm}$) case. a,b) Relative ionization degree $\rho$ at the end of the simulation with regions of overcritical plasma density indicated by white-dashed lines. c,d) Maximum local intensity reached during pulse propagation. (e,f) Snap-shot of the field amplitude at pulse peak.}
    \label{fig:fig1}
\end{figure}

Figure \ref{fig:fig1} shows an example for the shortest pulse duration and strongest focusing conditions considered in our analysis and highlights the key importance of the feedback effects within the description of the pulse propagation and plasma dynamics. It displays the final plasma density profile (top row) with the associated region of overcritical plasma density (white dashed contour line), the profile of the local peak intensity (middle row), and the instantaneous field profile sampled at the moment where the pulse peak passes the focal point.

Two main features are observed by the comparison of the plasma density and peak intensity maps. With full feedback included, both maps show substantially reduced peak values and a pronounced asymmetry with respect to the focal plane that is absent without feedback. The overestimation of the overall signal magnitude for the plasma density and peak intensity in the simulation without feedback is mainly attributed to the neglect of the field energy removal associated with plasma generation and heating. However, the pronounced asymmetry and overall more complex structure of the profiles with feedback can only be explained with transient non-trivial propagation effects. Specifically, the additional streaks in Fig. \ref{fig:fig1}(c) (see arrows) document the interplay of nonlinear absorption and propagation. Moreover, when compared to the reference in \ref{fig:fig1}f the field distribution with feedback in Fig. \ref{fig:fig1}(e) displays a complex structure with interference fringes and wave contributions from different phase front curvatures. Overall, the example displayed in Fig. \ref{fig:fig1} documents substantial impact of the plasma feedback on plasma generation, intensity profile as well as temporal and spatial field structure.

\begin{figure}
    \centering
        \small
    \includegraphics[width=0.70\linewidth]{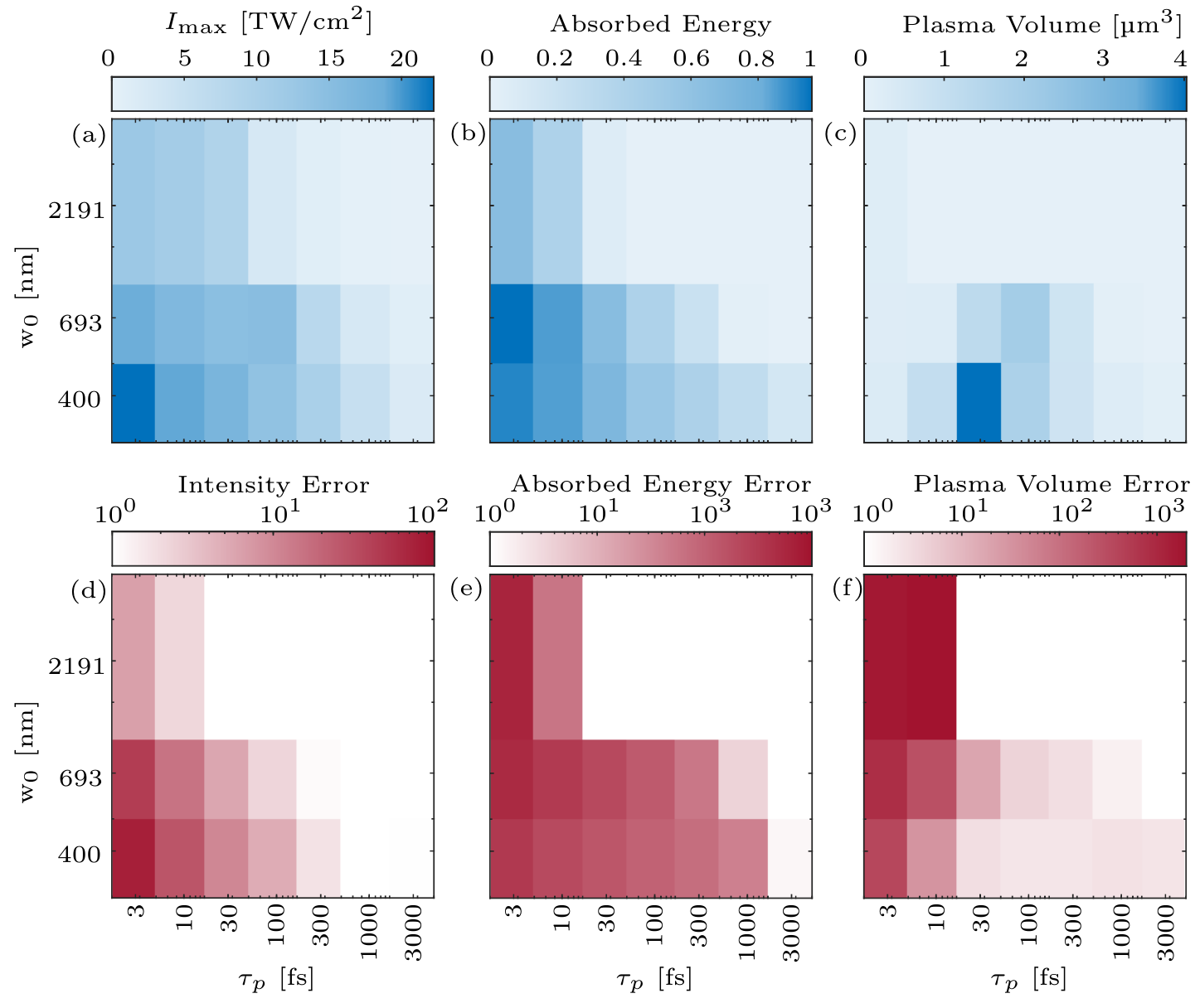}
      \normalsize
    \caption{Top row: Systematic study of (a) peak intensity, (b) total absorbed energy and (c) overcritical plasma volume as function of pulse duration and focus spot size for calculations including plasma feedback. All three observables show maxima at different beam conditions. Bottom row: Corresponding ratio of values without and with feedback included. White areas indicate parameter regions where feedback is of minor importance.}
    \label{fig:fig2}
\end{figure}

Before we perform a more detailed analysis of the individual features and their origin we explore the significance of the feedback as function of pulse duration and focus spot size based on three central observables. In particular, we analyze the overall peak intensity, the total absorbed energy as well as the volume with overcritical plasma density in the simulation arena. The corresponding maps for a description with full feedback are displayed in Fig.~\ref{fig:fig2}(a-c).

We begin with the discussion of the peak intensity. As intuitively expected, the highest peak intensity is found for the shortest pulse and strongest focusing conditions (3fs, 400nm). However, the intensity gain with decreasing spot size and/or decreasing pulse duration is much smaller than predicted without nonlinear plasma feedback. The impact of the latter is quantified by the ratio of results without feedback compared to those with feedback as shown in the map in Fig. \ref{fig:fig2}(d) for the whole parameter space. Specifically for the case with strongest focusing and shortest pulse duration a feedback-free description overestimates the peak intensity by a factor of more than an order of magnitude. For the weakest focusing and the longest pulse, however, the predictions of both simulations coincide, documenting that plasma effects are negligible at the associated low intensities. Overall, it can already be stated that in the lower left triangle of the parameter map a full account of the plasma feedback is mandatory for a quantitative description of the physics.

\begin{figure}
    \centering
    \small
    \includegraphics[width=0.50\linewidth]{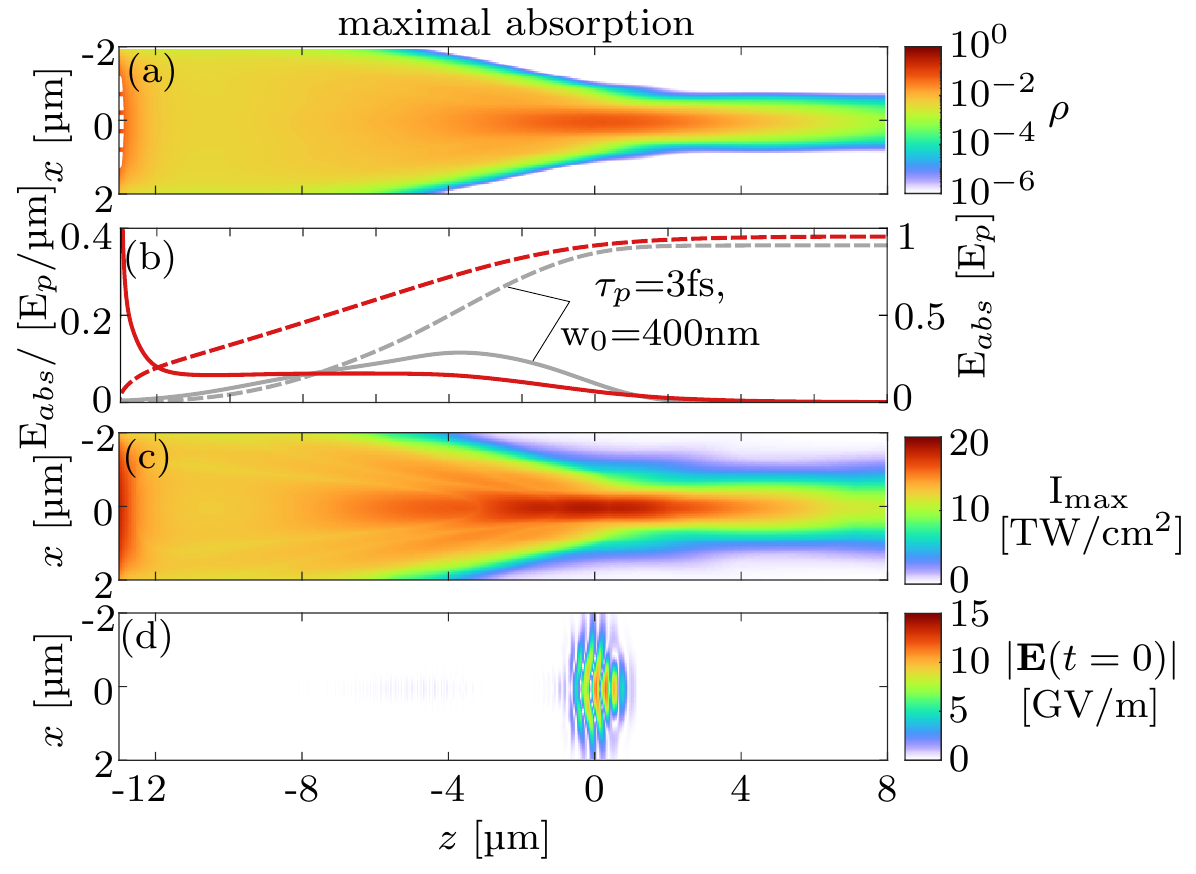}
    \normalsize
    \caption{Analysis of maximum absorption scenario (3fs, 400nm). a) Final plasma density map with overcritical regions indicated by white-dashed lines. b) Local (solid) and accumulated (dashed) energy deposition along the propagation axis. Results for the shortest pulse and strongest focusing case are shown in gray for comparison. c) Local peak intensities reached during pulse propagation. d) Electric field amplitude snapshot at pulse peak.}
    \label{fig:fig3}
\end{figure}
We proceed with a discussion of the total absorbed energy. Maximum absorption occurs for the shortest pulses and, surprisingly, not for strongest but intermediate focusing, see map in Fig.~\ref{fig:fig2}(b). The reason for this effect can be identified in the more detailed analysis of this scenario displayed in Fig.~\ref{fig:fig3}. While the peak intensities and plasma densities still have a local maximum near the focal spot (see panels ~\ref{fig:fig3}a,c), substantial ionization and therefore also energy absorption can be observed all the way from the focus region to the silica surface. In fact the global maximum of the plasma density is found in the plasma mirror emerging in the surface layer [cf. Fig. ~\ref{fig:fig3}(a)]. This is in stark contrast to the results found for strongest focusing (cf. Fig. \ref{fig:fig2}) where the global maximum is clearly inside the focus without a noticeable density increase near the surface. A particularly clear picture of the underlying physics can be deduced from the local and accumulated energy deposition along the propagation axis shown in fig. \ref{fig:fig3}(b). For the strongest focusing case (light gray lines) the local absorption peaks near the focus and decays rapidly before and after the peak due to the intensity drop of the highly divergent beam. For the case of slightly weaker focusing (693nm) the reduced peak intensity leads to significantly lower absorption in the focus region, compare solid lines around $z=0$. At the same time, due to the weaker divergence of the beam, intensities at the front surface are higher. In fact, the resulting additional ionization and heating overcompensate the reduced absorption in the focus region slightly, see accumulated energy deposition in Fig.  \ref{fig:fig3}(b). Further increasing the focus width eventually leads to peak intensities below the ionization threshold, entering the realm of linear pulse propagation.

\begin{figure}
    \centering
    \small
    \includegraphics[width=1.00\linewidth]{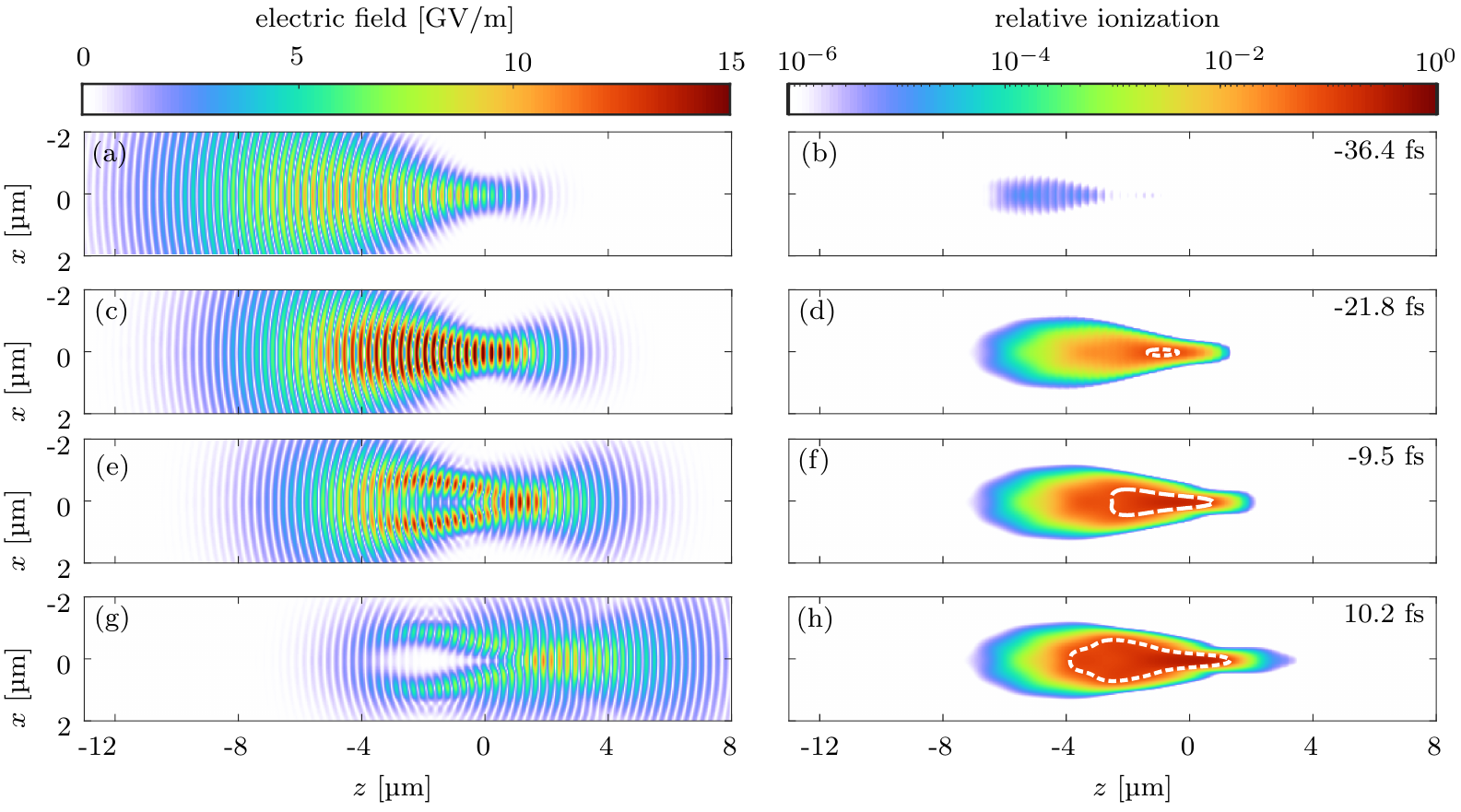}
    \normalsize
    \caption{Electric field evolution (left) and plasma generation dynamics (right) for the maximum critical plasma generation scenario (30fs, 400nm). The increasing overcritical plasma volume forces electromagnetic fields to propagate around these regions, leading to a transient hollow field profile in the focus.}
    \label{fig:fig4}
\end{figure}
Finally, in Fig.~\ref{fig:fig2}(c), we examine the plasma generation efficiency, which exhibits another surprising behavior: the maximum critical plasma volume is achieved not for the shortest pulse, but for 30~fs pulse duration (tightest focusing). To get a more detailed understanding of the underlying dynamics we analyze the electric field evolution and plasma generation dynamics shown in Fig.~\ref{fig:fig4}. Ionization starts a few $\mu m$ in front of the focus spot, where the threshold intensity is exceeded first (see panels \ref{fig:fig4}a,b). Critical plasma density, however, is reached first around the focal spot area (see panels c, d).  After the front of the pulse creates this ionized overcritical region, the center of the pulse diffracts from it and propagates around it, leading to the hollow field profile seen in Fig. \ref{fig:fig4}(e). At this point the field is still sufficiently strong to ionize the medium in its path, mainly via avalanching in this already photo-activated region, thereby extending the region of overcritical plasma layer by layer. In contrast, for shorter pulses as shown in Fig. \ref{fig:fig1}, the intensity in the focus is very high, and a small spot of very dense plasma is produced. The remaining part of the pulse is, however, too short to drive a significant electron impact avalanche to extend this region. For longer pulses, on the other hand, the photo-activated volume that acts as a precursor for the avalanching is getting smaller and eventually missing completely. An additional interesting feature visible in Fig. 3(h) is the appearance of the secondary focus which is formed by the radiating which circumvented the overcritical plasma region and recollimated at around $z=2$ $\mu$m. This effect can be compared to the spatial replenishment during filamentation in air \cite{repl}.

\section{Conclusion}
We have established a comprehensive first-principle model for the high-intensity pulse propagation in solid dielectrics. The model includes all effects relevant for ultrashort pulse propagation with intensities up to 10$^{15}$ W/cm$^2$ and above, such as photoionization, avalanche ionization, change of the electron velocity distribution, etc. The model allows accurate treatment of the plasma densities above critical and arbitrary focusing geometry due to the FDTD formulation. The model was tested for several cases of tightly focused pulse propagation in fused silica. Surprisingly, we found that maximum energy deposition as well as maximum critical plasma generation do not occur for the shortest pulses and tightest focusing conditions, illustrating the importance of the dynamical feedback between transient optical properties and plasma dynamics for the physically correct description of the associated laser-matter interaction.

\section{Acknowledgement}
We acknowledge fruitful discussions with Timm Bredtmann \tcb{and his contribution to developing the concept of the code}. This work was supported by the United States Air Force Office of Scientific Research under program No. FA9550-12-1-0482. We acknowledge financial support from the German Research Foundation (DFG) via CRC 1477 “Light-Matter Interactions at Interfaces” (ID: 441234705).

\end{document}